# WDM/TDM over Passive Optical Networks with Cascaded-AWGRs for Data Centers

Mohammed Alharthi[1], Sanaa H. Mohamed[2], Taisir E. H. El-Gorashi[2], Jaafar M. H. Elmirghani[2]

[1]School of Electronic and Electrical Engineering, University of Leeds, LS2 9JT, United Kingdom
[2]Department of Engineering, King's College London, WC2R 2LS, United Kingdom
elmaalh@leeds.ac.uk, s.h.h.mohamed@leeds.ac.uk, taisir.elgorashi@kcl.ac.uk, jaafar.elmirghani@kcl.ac.uk

**ABSTRACT**
Data centers based on Passive Optical Networks (PONs) can provide high capacity, low cost, scalability, elasticity and high energy-efficiency. This paper introduces the use of WDM-TDM multiple access in a PON-based data center that offers multipath routing via two-tier cascaded Arrayed Waveguide Grating Routers (AWGRs) to improve the utilization of resources. A Mixed Integer Linear Programming (MILP) model is developed to optimize resource allocation while considering multipath routing. The results show that all-to-all connectivity is achieved in the architecture through the use of two different wavelength within different time slots for the communication between racks in the same or different cells, as well as with the OLT switches.

## 1. INTRODUCTION

The traffic volumes in need of processing and transporting have massively increased in recent years due to the growth in using Internet- based applications [1]. Research efforts have optimized the designs of access and core communication networks [2]-[13] as well as the design of data centers [13]-[24] to meet the requirements of the increasing Internet traffic while maintaining energy-efficiency. Among other limitations and challenges facing current data center architectures are the high cost, high latency, low throughput, management complexity, and limited scalability [25]-[27].

A new trend of research is focusing on introducing Passive Optical Network (PON) technology in data center networks [28]-[31]. The results in these studies discussed the ability of PON technology to provide high capacity, low cost, elasticity, scalability, and energy efficiency for future data centers. Passive devices such as, Passive Polymer Backplane, Fibre Bragg Grating (FBG) and passive star reflector are used in these architecture to maintain the communication between servers in the same rack [17].

The Wavelength Division Multiplexing (WDM) PON, Orthogonal Frequency Division Multiplexing (OFDM) PON are used in PON-based data center networks [29]-[32]. An WDM AWGR-based PON data center architecture was introduced in [24], and [33]. The wavelength assignment and routing for inter-rack communication is optimised in the WDM architecture to achieve all-to-all connectivity while considering a single path between source and destination pairs. The work in [5] introduced WDM-TDM in the architecture proposed in [24] by considering sharing the wavelengths among source and destination pairs by allocating time slots.

This paper proposes the use of WDM-TDM technique in a two-tier cascaded-AWGRs data center architecture that considers multipath routing which is discussed in [34], [35]. Using WDM-TDM in the multipath two tier cascaded-AWGRs data center architecture enables building a more efficient architecture by dividing the available wavelength resources, into several time slots. This enables fine-granular resources assignment based on wavelength and time slots for the communication between the racks in different cells and between racks and the OLT switches.

The reminder of this paper is organized as follows: Section 2 describes the use of WDM-TDM multiple access over the PON-based data center with cascaded-AWGRs. Section 3 describes the MILP optimization model. Section 4 presents and discusses the results. Section 5 provides the conclusions of this paper.

## 2. THE USE OF WDM/TDM OVER THE PON-BASED DATA CENTERS WITH CASCADED-AWGRS

The two tier cascaded AWGRs in the PON-based data center architecture in [34] provides multipath passive connectivity between entities within the data center. Figure 1 illustrates the design of the PON-based data center architecture with two tier cascaded AWGRs. In Figure 1, the architecture consists of four cells and each cell has four racks that are interconnected through a special server. Considering WDM, the number of required wavelengths is equal to 2N [34], where N refers to the number of cells and OLT switches and the the size of each AWGR equals N×N. If intra-cell communication is not required via the two tier cascaded AWGRs, the number of wavelengths equals 2(N-1). For instance, the proposed architecture as depicted in Figure 1 includes 4 cells and 4 OLT switches which means N=8. Therefore, the number of wavelengths is equal to 16 when considering intra and inter cell communications. When only inter cell communications is required, the number of wavelengths is equal to 2(8-1)=14. Intra rack communication can be achieved by employing a passive connection according to [3].

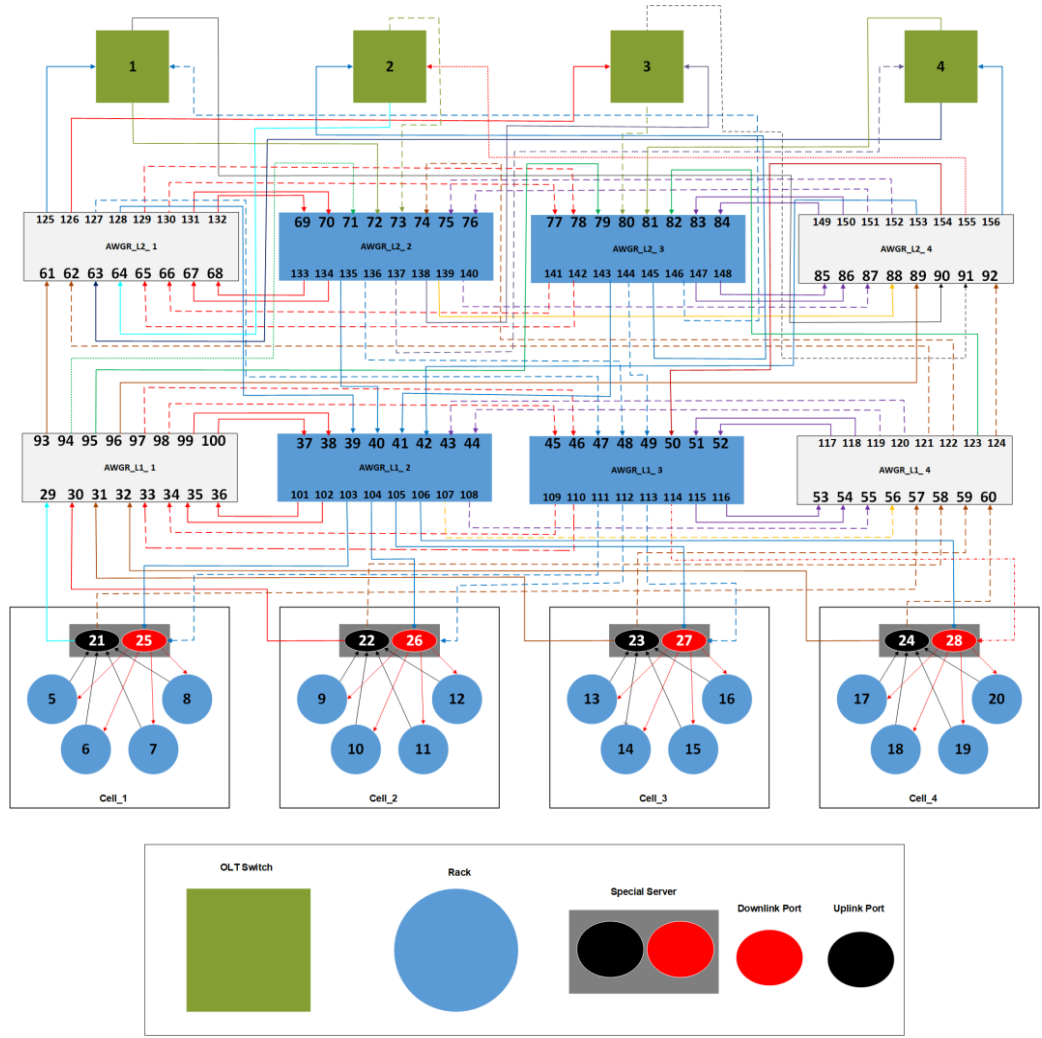

*Figure 1.The two tier cascaded AWGRs architecture with four cells.*

When considering WDM-TDM, the communication between the racks in the same cell or different cells as well as between racks and OLT switches can be assigned using several wavelengths and sent over different time slots. The special servers include a database that contains addresses of servers and the wavelength and time slot allocated to each rack. The special servers communicate with each other through OLT switches. Each special server has two links for uplink and two links for downlink, each is connected to level one of the two-tier cascaded AWGRs. The two-tier cascaded AWGRs routes traffic provides two paths to route traffic. The alternative path provides load balancing at high traffic load and resilience.

The special server receives requests from servers and if it decides to grant the request, it replies with control messages to the servers that contain information about the wavelength and time slot tuning required. The two tier cascaded AWGRs route the source server's traffic using the assigned route and wavelength through the two tier cascaded AWGRs until it reaches the receiving server. The special servers communicate with the OLT switches to exchange information and update their databases.

## 3. TDM-WDM MILP MODEL OPTIMIZATION

In this section, we briefly describe the Mixed Integer Linear Programming (MILP) model we developed to optimize the static allocation of wavelengths and time slots in the two tier cascaded AWGR-based PON data center architecture. We consider an architecture that contains two cells and two OLTs due to the complexity of running MILP model at larger scale. Each cell has two racks, and the racks within a cell are connected to a special server. Two layers each with two 4×4 AWGRs are used to connect the two cells and the two OLT switches. The number of wavelengths needed in this case is eight wavelengths to achieve multipath routing. The MILP model aims to maximize the total number of connections between the racks as well as between racks and the OLT switches as indicated in Equation 1. $\gamma_{sd}^{jt}$ is a binary variable that is equal to 1 if source s, $s \in G$, where $G$ is the set of all communicating entities (i.e., OLT switches and racks), and destination d, $d \in G$ are assigned to time slot t, $t \in T$ where $T$ is the set of all time slots, and wavelength j, $j \in W$, where $W$ is the set of all wavelengths. The optimization model considers all routing restrictions and wavelength allocation constraints to achieve the objective for the proposed architecture.

$$\textit{Maximize:} \sum_{\substack{s \in G \\ s \neq d}} \sum_{d \in G} \sum_{j \in W} \sum_{t \in T} \gamma_{sd}^{jt}, \quad (1)$$

## 4. RESULTS AND DISCUSSIONS

The work in [34] presented a MILP model to optimize the routing and wavelength assignment for passive PON with 2-tier cascaded AWGRs connecting 4 cells and 4 OLT switches while considering multipath routing. The results indicated achieving all-to-all inter-cell/OLT communication in multipath style. The assigned wavelength should be shared among the servers within the cell which can lead to under-utilization of resources and increased latency due to blocking other servers from using that wavelength.

This work introduces the utilization of TDM besides WDM over the passive optical networks with cascaded-AWGRs for data centers to ensure that each rack can use a wavelength in a certain time slot. The WDM-TDM technology enables the racks to send/receive to/from a specific rack/OLT switch by utilizing the assigned wavelengths and time slots. Table 2 demonstrates the MILP model results of the wavelength and time resource assignment. Each rack/OLT switch uses two different wavelength to communicate with other racks and the OLT switches to achieve multipath routing [34] as shown in Table 2. For example, Rack 1 in cell 1 can communicate with OLT switch 1 by using either wavelength 3 in time slot 5 or wavelength 7 in time slot 2.

*Table 2: MILP-BASED RESULTS FOR RESOURCE ASSIGNMENT IN THE PON WITH 2-TIER CASCADED AWGRS. EACH PAIR ARE ASSIGNED TWO DIFFERENT WAVELENGTHS AND TIME SLOTS FOR THE COMMUNICATION.*

| | | Cell 1 | | Cell 2 | | OLT_1 5 | OLT_2 6 |
| --- | --- | --- | --- | --- | --- | --- | --- |
| | | Rack_1 71 | Rack_2 72 | Rack_3 73 | Rack_4 74 | | |
| Cell 1 | Rack_1 71 | | λ5, τ6 / λ8, τ10 | λ1, τ4 / λ6, τ9 | λ1, τ8 / λ6, τ5 | λ3, τ2 / λ7, τ1 | λ2, τ3 / λ4, τ7 |
| | Rack_2 72 | λ5, τ2 / λ8, τ8 | | λ1, τ7 / λ6, τ3 | λ1, τ9 / λ6, τ1 | λ3, τ4 / λ7, τ5 | λ2, τ10 / λ4, τ6 |
| Cell 2 | Rack_3 73 | λ3, τ10 / λ7, τ9 | λ3, τ1 / λ7, τ5 | | λ4, τ3 / λ5, τ4 | λ2, τ7 / λ8, τ6 | λ1, τ8 / λ6, τ2 |
| | Rack_4 74 | λ3, τ7 / λ7, τ1 | λ3, τ2 / λ7, τ9 | λ4, τ10 / λ5, τ6 | | λ2, τ3 / λ8, τ8 | λ1, τ4 / λ6, τ5 |
| OLT_1 5 | | λ4, τ3 / λ6, τ6 | λ4, τ4 / λ6, τ7 | λ2, τ5 / λ7, τ8 | λ2, τ2 / λ7, τ10 | | |
| OLT_2 6 | | λ1, τ4 / λ2, τ5 | λ1, τ8 / λ2, τ3 | λ3, τ2 / λ8, τ1 | λ3, τ7 / λ8, τ6 | | |

## 5. CONCLUSIONS

This paper introduced the use of WDM-TDM in a two tier cascaded AWGRs data center that support multipath routing and presented the results of a MILP model that optimizes the allocation of wavelengths and time slots. The results showed that all-to-all multipath connectivity is achieved which means that each rack within each cell can use two different wavelengths, each in a certain time slot to communicate with other destinations. The use of WDM-TDM with multipath routing can lead to improvements in the resource utilization as well as it can resolve some of the most common issues in data centers including congestion, blockage and oversubscription.

**ACKNOWLEDGEMENTS**
The authors would like to acknowledge funding from the Engineering and Physical Sciences Research Council (EPSRC), INTERNET (EP/H040536/1), STAR (EP/K016873/1) and TOWS (EP/S016570/1) projects. All data are provided in full in the results section of this paper. The first author would like to thank the Ministry of Interior (MOI), Saudi Arabia for funding his PhD scholarship.